\documentclass[11pt]{amsart}
\usepackage{amssymb}
\usepackage{amsmath}
\usepackage{amsthm}
\usepackage{graphicx}
\usepackage{hyperref}
\numberwithin{equation}{section}

\theoremstyle{remark}

\newcommand{\bA}{\mathbf{A}}

\newcommand{\bB}{\mathbf{B}}
\newcommand{\br}{\mathbf{r}}
\newcommand{\bn}{\mathbf{n}}
\newcommand{\bzero}{\mathbf{0}}

\newcommand{\bZ}{\mathbb{Z}}
\newcommand{\nab}{\mathbf{\nabla}}
\begin{document}
\title[ M\"obius band flux quantization]{Flux quantization for a\\
  superconducting ring\\ 
in the shape of a M\"obius band}
\author{Jonathan Rosenberg}
\thanks{JR partially supported by NSF Grants DMS-0805003
and DMS-1206159.}
\address{Department of Mathematics\\
University of Maryland\\
College Park, MD 20742, USA}
\email{jmr@math.umd.edu}
\urladdr{http://www.math.umd.edu/\raisebox{-.6ex}{\symbol{"7E}}jmr}
\author{Yehoshua Dan Agassi}
\address{Naval Surface Warfare Center, Carderock Division\\
9500 MacArthur Blvd.\\
West Bethesda, MD 20817, USA}
\email{yehoshua.agassi@navy.mil}
\keywords{M\"obius band, superconductor, flux quantization, $U(1)$-gauge
  bundle} 
\subjclass[2010]{Primary 82D55; Secondary 53C05}
\begin{abstract}
We give two derivations of magnetic flux quantization in a
superconducting ring in the shape of a M\"obius band, one using direct
study of the Schr\"odinger equation and the other using the holonomy
of flat $U(1)$-gauge bundles. Both methods show that the magnetic flux
must be quantized in integral or half-integral multiples of
$\Phi_0=hc/(2e)$. Half-integral quantization shows up in ``nodal
states'' whose wavefunction vanishes along the center of the ring, for
which there is now some experimental evidence.
\end{abstract}
\maketitle

\begin{figure}[hbt]
\begin{center}
\includegraphics*[width=3in,clip]{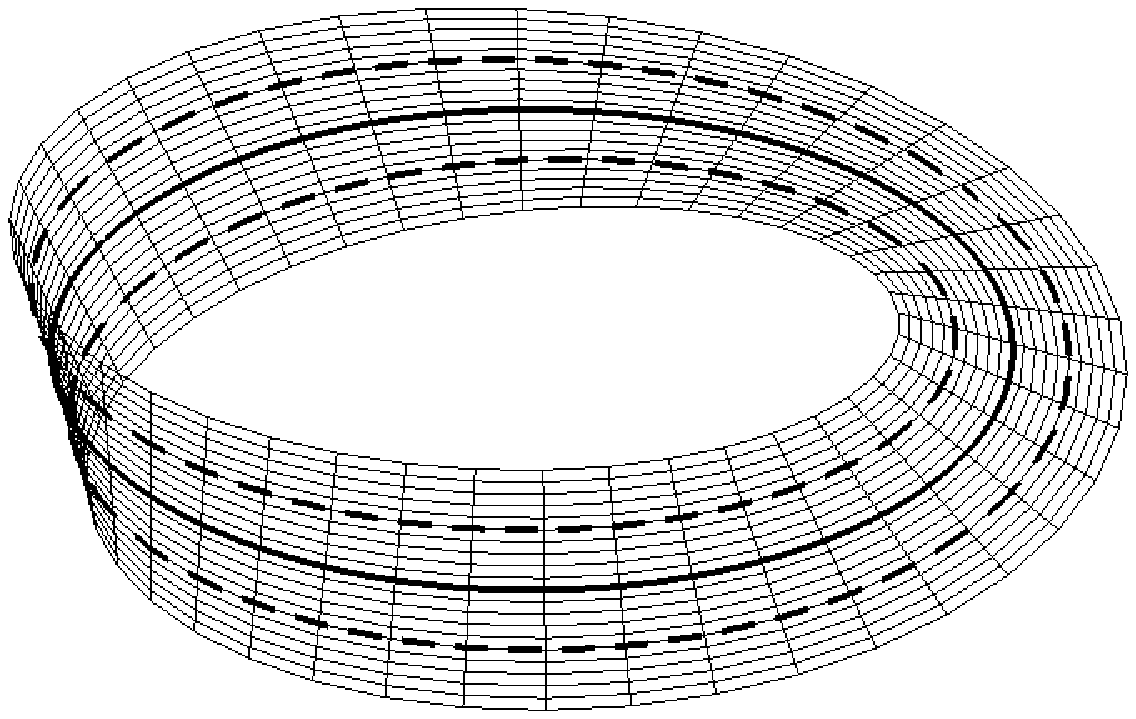}

\includegraphics*[width=3in]{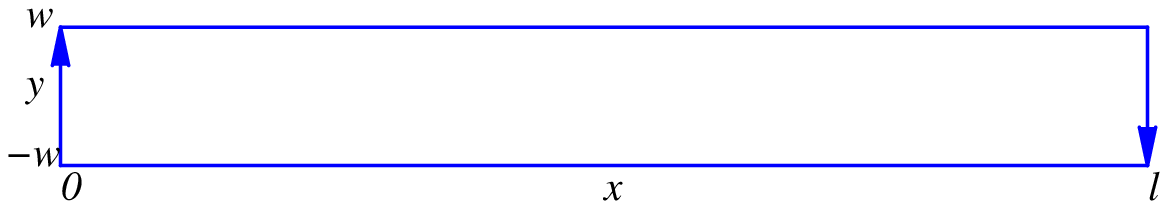}
\end{center}
\caption{A M\"obius band with its coordinate system}
\label{fig:mob}
\end{figure}

One of the best-known macroscopically observable quantum effects is
the quantization of magnetic flux through a superconducting ring $M$
in units of $\Phi_0 = \frac{hc}{2e}$ \cite{DvrFbnk}. The question we
treat here is this: 
What happens to this condition if the superconducting ring $M$ is in the
shape of a M\"obius band (Figure \ref{fig:mob}) rather than an
annulus? Does the condition 
remain the same, or should it be modified? This provides a good test
case for the application of topology in physics. It should be possible
to check the result experimentally since crystals of potentially
superconducting materials such as NbSe$_3$
have recently been produced with a M\"obius band
shape \cite{MobCrystal,MobCrystal1}. Investigation of a different
topology, the ``figure eight,'' was carried out by Vodolazov and
Peeters \cite{Peeters}.

Several possible derivations for flux quantization have been
given. We shall analyze what happens to them under this ``exotic''
topology, under the assumption that the superconductor is extremely
thin and can be treated as if it were $2$-dimensional. 
We can take coordinates in the superconductor to be $x$ and
$y$, with $x$ (going around the ring $M$) going from $0$ to $\ell$, $y$
going from $-w$ to $w$, and the point $(0,y)$ identified with $(\ell, -y)$.
The simplest treatment (following \cite{McInnes},
based roughly on the classic paper \cite{ByersYang})
starts with the Schr\"odinger equation for a
Cooper pair of charge $-2e$ in an electromagnetic field with circular
symmetry. Recall that the vector potential $\bA$ is not necessarily
globally defined on the ring, though we can take it to be single-valued
on $[0,\ell]\times [-w, w]$, possibly with different values at $(0,y)$
and at $(\ell,-y)$. 
By the Meissner effect, $\bB=\nab\times \bA=\bzero$ in the
superconductor $M$, which implies that $\bA$ is \emph{locally} of the form
$\nab\phi$. Again we can take $\phi$ to be single-valued on
$[0,\ell]\times [-w, w]$, possibly with different values at $(0,y)$ 
and at $(\ell,-y)$. The (time-independent) Schr\"odinger equation becomes
\begin{equation}
\label{eq:Sch}
\frac{1}{2m}\bigl( -i\hbar\nab + \frac{2e}{c}\bA \bigr)^2 \psi + V\psi
= E\psi,
\end{equation}
where $\psi$ is the wavefunction of a Cooper pair, treated a single
boson with mass $m$. Let $\psi_0$ be a solution of \eqref{eq:Sch} with
$\bA=\bzero$. Then if $\psi=\psi_0e^{-i\alpha\phi}$ for a constant
$\alpha$, we have
\[
\begin{aligned}
\bigl( -i\hbar\nab + \frac{2e}{c}\bA \bigr)\psi &=
\bigl( -i\hbar\nab\psi_0 \bigr)e^{-i\alpha\phi} 
-\hbar\alpha \psi_0e^{-i\alpha\phi}\bA
+\frac{2e}{c}\psi_0e^{-i\alpha\phi}\bA\\ 
&=
-i\hbar e^{-i\alpha\phi}\nab\psi_0,
\end{aligned}
\]
provided that $\alpha=\frac{2e}{\hbar c}$. Thus $\psi$ will be a
solution of \eqref{eq:Sch} for this value of $\alpha$. Since the
wavefunction must be well-defined globally, in a ground state where
$\psi_0$ is everywhere non-zero on the interior of $M$, 
$\phi(\ell,-y)-\phi(0,y)$
must have a constant value of $\frac{2\pi n\hbar c}{2e}= n\frac{h
  c}{2e} = n\Phi_0$, for some $n\in \bZ$, independent of $y$. In
particular, if $C$ is the closed curve through the middle of the
superconductor, corresponding to $y=0$ in our coordinate system (see
solid black curve in Figure \ref{fig:mob}), then $\oint_C \bA\cdot d\br =
n\Phi_0$, which is the flux quantization condition. Note however that
if we take the curve $C'$ given by $y=\text{const.}$ with the constant
non-zero, then when $x$ runs from $0$ to $\ell$, we end up on the opposite
side of $C$ from where we started, and so one has to let $x$ run all
the way out to $2\ell$ to get back to the starting point (see dashed black
curve in Figure \ref{fig:mob}). Thus for $C'$, we have the modified
quantization condition $\oint_{C'} \bA\cdot d\br =
2n\Phi_0$, with an extra factor of $2$.

However, another phenomenon, proposed on different theoretical grounds
(calculations with the Ginzburg-Landau and Bogoliubov-de Gennes models)
in \cite{Hayashi1,Hayashi}, and supported by experimental evidence in
\cite{Hayashi2}, is also possible. Namely, one can have a ``nodal''
state supported near the dashed curve $C'$, with $\psi_0=0$ on the
circle $C$ given by $y=0$. Then it suffices to for
$\phi(\ell,y)-\phi(0,y)$ to be a \emph{half-integer} multiple of
$\Phi_0$. This will still give a globally single-valued wavefunction
since the point $(\ell,y)$ is identified with $\phi(0,-y)$ and 
$\phi(\ell,-y)-\phi(0,-y)$ is again a half-integer. Thus the flux
appears to be quantized in half-integral units of $\Phi_0$, as was
observed in \cite{Hayashi2}. A slightly different analysis by Mila
\emph{et al.} 
\cite{Mila} showed that for a M\"obius ``ladder'' geometry,
quantization should be in multiples of $\Phi_0$ ``as long as coherent
motion between the chains is possible,'' and in multiples of
$\Phi_0/2$ when there is no coherent motion between the chains (so
that particles effectively travel along the curve $C'$).

A more sophisticated approach follows the ideas of \cite{McInnes}. 
Following Dirac \cite{Dirac} and Weyl, we view the magnetic field as a
$U(1)$-gauge field. More precisely, the vector potential $\bA$ is a
connection on a $U(1)$-bundle and the field strength (curvature
$2$-form) is the magnetic field. The \emph{phase} of the wave function
$\psi$ is a section 
of this $U(1)$-bundle \cite{Dirac}. Because of the Meissner effect, the
curvature of the bundle vanishes on $M$, i.e., the bundle is flat. 
Since $H^2(M,\bZ)=0$, the bundle is also topologically trivial, and
the connection differs from the usual connection (corresponding to the
case $\bA=\bzero$) by a constant $1$-form, which we can identify with
a single real number, which is the value of the holonomy or flux
$\Phi=\oint_C \bA\cdot d\br$. Since the wavefunction must be single-valued
on $M$, and since as we saw above, the change in the wavefunction as
we go around the loop $C$ is 
\[
e^{-i\frac{2e}{\hbar c}\Phi},
\]
we obtain the flux quantization condition $\Phi=n\Phi_0$.

The ``nodal state'' case can be treated similarly, except that we
replace $M$ by the complement of $C$ since we are assuming the Cooper
pairs are localized away from the center of the M\"obius band. (If the
wave function vanishes on $C$, then its phase there is not
well-defined, so the bundle is not defined on $C$.) 
Since the inclusion $(M\smallsetminus C)\hookrightarrow M$ sends a
generator of the fundamental group of  $M\smallsetminus C$ to
\emph{twice} a generator of the fundamental group of  $M$, we have in
effect twice as many possible flat bundles. In other words, we only
require $\oint_{C'} \bA\cdot d\br$ to be integral, which means that
flux around the superconducting loop satisfies 
the flux quantization condition $\Phi=n\Phi_0$ with $n$ a half-integer.

The second approach also explains the answer to another question: the
flux $\Phi=\oint_C \bA\cdot d\br$ only depends on the homology class
of the loop $C$ in $H_1(M,\bZ)\cong \pi_1(M)\cong \bZ$, since this is
a basic fact about holonomy of flat connections. In particular,
since $C'$ is homologous to $2C$, the flux around $C'$ is quantized in
twice the units of the flux around $C$.
From the first point of view, this is a bit harder to
see, since if $C_1$ and $C_2$ are homologous loops, the usual argument
(when $M$ is an annulus) would be to take the region $D$ with boundary
$C_1-C_2$ (i.e., $C_1\cup C_2$, but with reversed orientation on
$C_2$) and to use Stokes' Theorem to argue that
\begin{equation}
\label{eq:Stokes}
0 = \iint_D (\nab\times \bA)\cdot \bn = \oint_{C_1} \bA\cdot d\br -
\oint_{C_2} \bA\cdot d\br.
\end{equation}
This runs into difficulties since $M$ is not orientable (so that the
normal $\bn$ is not well defined), and thus
Stokes' Theorem  doesn't seem to apply. But one can rectify things by
cutting $M$ open along a curve whose complement is orientable.

An issue not settled by our analysis is what happens with a ``thick''
M\"obius band (thickness being measured compared to the penetration
depth), where $2$-dimensionality is no longer a good assumption. 
In this case, since there may not be nodal states in the sense described
above, one might expect flux quantization only in integral multiples
of $\Phi_0$. It would be interesting to test for this experimentally,
as the results of \cite{Hayashi2} are only for a thin film.

\end{document}